\documentclass{elsart}
\usepackage{latexsym}
\usepackage{epsfig}
\usepackage{natbib}

\def\url#1{#1} 

\begin{document}

\begin{frontmatter}

\title{The incompatibilities between the standard theory of interstellar 
extinction and observations.}

\author{Fr\'ed\'eric \snm Zagury}
\address{
\cty 02210 Saint R\'emy Blanzy, \cny France
\thanksref{email} }

   \thanks[email]{E-mail: fzagury@wanadoo.fr}

\accepted{February 2002}

 \begin{abstract}
In this paper I review a series of observations which do not agree with 
the standard interpretation of the extinction curve. 
The consequence is that light we receive from a reddened star must be contaminated 
by starlight scattered at very small angular distances from the star. 
The true extinction curve is a straight line from the near infrared to the far-UV. 
If so, all interstellar grains models must be questionned. 
Another conclusion concerns the average properties of interstellar grains 
which seem much more uniform than previously thought.
\end{abstract} 
\end{frontmatter}
 \section{Introduction} \label{intro}
Although interstellar dust represents a very small part of interstellar matter, 
it is responsible for the extinction of starlight, 
it accounts for a large part of the scattered starlight we receive from nebulae, 
and it is also responsible for the thermal emission of interstellar clouds in the infrared. 
Thus, interstellar dust plays a most important role as a tracer of the 
distribution and of the structure of interstellar matter. 
It is also involved in the chemistry of the interstellar 
medium.

The study of interstellar dust has led to several dust models 
that are heavily constrained by the average extinction curve obtained from 
observations of stars through interstellar clouds. 
It is of primary importance to have a correct interpretation of the extinction curve
to fully understand the interstellar medium.

The standard interpretation of the extinction curve separates the 
light spectrum into three parts: 
the visible, the $2000\rm\AA$ bump, and the far-UV region. 
Separate types of particles (grains or molecules) with specific size 
distributions and extinction properties are assumed 
to be responsible for the extinction of light in each of these wavelength domains. 
The variation of the extinction curve with the line of sight is interpreted as the effect of 
different proportions of each type of particles from cloud to cloud. 
All current dust models are based on this paradigm.

In a preceding series of papers \citep{uv1,uv2,uv3,uv5,uv6} I have detailed 
observations which contradict these models, question the standard 
interpretation of the extinction curve and call for another explanation. 
There are important consequences involved, as we need to reconsider the nature of 
light received from a reddened star. Which, in turn, will deeply affect
the analysis and the interpretation of the observations on interstellar matter.
The applications are numerous ranging from practical aspects 
(correcting the reddening for stellar distance estimations), 
to more theoretical problems on the nature of interstellar dust, its properties 
depending on environment, etc...

In this paper I first reviewed the principles of the standard interpretation of the extinction curve
(section~\ref{standard}). 
The observations which run contradictivelly this interpretation are summarised 
in  section~\ref{obs}.
The implications and ways to reconcile theory and observation are 
discussed in section~\ref{conc}.
\section{The extinction curve and its' standard interpretation} 
\label{standard}
The light we receive from a reddened star is extinguished by a factor 
$e^{-\tau_{\lambda}}$, where $\tau_{\lambda}$ is the extinction 
optical depth, at wavelength $\lambda$, of the interstellar matter 
between the star and us.
If $F_{\lambda}$ and $F_{0\lambda}$ are the flux we receive from the 
star and the one we would receive if the star was not reddened:
\begin{equation}
    F_{\lambda}=F_{0\lambda}e^{-\tau_{\lambda}}
    \label{eq:extf}
\end{equation}¥
In magnitudes:
\begin{equation}
    m_{\lambda}=m_{0\lambda}+1.07\tau_{\lambda}=m_{0\lambda}+A_{\lambda}
    \label{eq:extm}
\end{equation}¥
Since $F_{0\lambda}$ or  $m_{0\lambda}$ are not known, they are 
replaced by the values observed for a non-reddened star  of same 
spectral type.
The extinction curve  $A_{\lambda}$ is then obtained, to within an 
additive constant generally determined from the $V$-magnitudes of the stars.
The extinction curve can be normalized by $E(B-V)=A_{B}-A_{V}$, 
proportional to the slope of the extinction curve in the visible.

\citet{seaton79} gave the average normalized extinction curve for the stars of the 
solar neighborhood.
The standard theory separates this curve into three parts: the 
visible, the $2200\rm\AA$ bump region, and the far-UV.
The normalised extinction curve in the direction of a reddened star 
follows Seaton's curve in the visible, but large variations are 
observed in the UV, especially in the far-UV \citep{bless72}.
The standard theory attributes these variations to the combined effect 
of the extinction of starlight by three types of particles, 
which are present, but in variable proportions, in the interstellar clouds.
The linear in $1/\lambda$ visible extinction is due to a distribution 
of large grains, which have a flat scattering cross section in the UV 
(figure~\ref{fig:seaton} in this paper or figure~2 in \citet{greenberg}).
The bump region is attributed to very small grains (VSG).
Last, the far-UV rise of the extinction curve should be due to 
molecules, thought to be poly-aromatic (the PAH).
`Large' or `small' particles refer to the wavelength domain which is 
considered, since the ratio of the size of the particle to the 
wavelength is the fundamental parameter in scattering theory.

By allowing the proportion of each type of 
particles to vary from cloud to cloud the standard theory acquires three 
degrees of freedom which permits to fit most extinction curves.
But, there does not seem to be any logic behind the grain type repartition 
with environment (density, exposure to UV radiation\ldots) 
\citep{jenniskens93}.

This freedom is paid with an important compensation.
The particular extinction curve each type of grains must have, the 
necessity to respect cosmic abundances, tightly constrain the nature 
of the grains and led to several models of grains.
To date, none of the models of interstellar dust in use (the first PAH model of \citet{desert}, the 
unified model of \citet{li97}, the model of \citet{mathis}) are truly satisfying.
\section{The standard theory against observations} 
\label{obs}
The standard theory can be tested in several ways.
I have proposed four tests, three of which are detailed in the 
following sub-sections.
\subsection{The UV spectrum of nebulae} 
\label{neb}
The UV spectrum of a nebula illuminated by a nearby star 
is well reproduced by 
the product of the spectrum of the source star and of a linear function 
of $1/\lambda$ \citep{uv1}. 

Neither the large grains supposed to be responsible for visible 
extinction -of which the far UV extinction cross section is nearly 
independent of wavelength-, 
nor the small particles supposed to be responsible for the UV extinction 
-which should scatter starlight as $1/\lambda^4$- can explain this 
relation between the nebula and the star spectra.

The linearity in $1/\lambda$ of the nebula to the star spectrum ratio
suggests that the scattering law valid in the visible extends to the UV. 

We also do not observe any excess of scattering in the bump region \citep{uv1}, 
which means that if there is a specific extinction at $2200 \,\rm\AA$, it is absorption only. 
But, some nebulae, associated to low-reddened stars, do not show a bump. 
Therefore, either the small grains which, according to the standard 
theory, are responsible for the bump are not present (or only in 
very small quantities) in low column density clouds, or the bump is not a common absorption process.
\subsection{The extinction curve in directions of very low reddening} 
\label{vlr}
The spectra of stars of same spectral type and very low reddening, 
not reddened enough to have a $2200 \rm\AA$ bump, 
differ one from the other by the same exponential of $1/\lambda$ in the 
visible and in the UV \citep{uv5}. 
Thus, the extinction law in very low column density media is the same in 
the visible and in the UV, 
which confirms what is already suggested by the UV observation of the nebulae.
\subsection{The extinction curve in directions of low reddening} 
\label{lr}
Increasing the reddening of the stars, the extinction law deviates 
from the $1/\lambda$ linear extinction in the far-UV first, 
whereas the linear visible extinction law still extends to the bump 
region \citep{uv2}. 
The reduced spectrum of the stars is an exponential in the near-UV.
This exponential prolongs in the UV the visible extinction of the 
stars' light.
The deviation from the visible extinction of the far-UV reduced spectrum clearly appears as an 
additional component superimposed on the tail of the exponential 
(figure~\ref{fig:hd62542}). 

Here again, the standard theory can not explain in a natural way the extension 
of the visible extinction law to the near-UV.
\section{Discussion} \label{conc}
The observations mentionned in section~\ref{obs} contradict the standard 
theory of interstellar extinction.

The only alternative to the standard theory was first mentionned by \citet{savage75}.
It implies that when a reddened star is observed in the UV a non negligible proportion of scattered light is 
re-introduced into the beam of the observation.

\citet{snow75} have rejected this hypothesis from UV observations of 
$\sigma$-Sco with two different apertures of $8'\times 3^\circ$ and $0.3''\times 39''$.
There is no difference between the UV spectra of the star observed 
with one aperture or the other.
However, these observations only prove that if scattered starlight contaminates the spectrum of reddened stars, 
it must be within an angle of $0.3''$ from the star. 

\begin{figure*}
\resizebox{!}{0.8\columnwidth}{\includegraphics{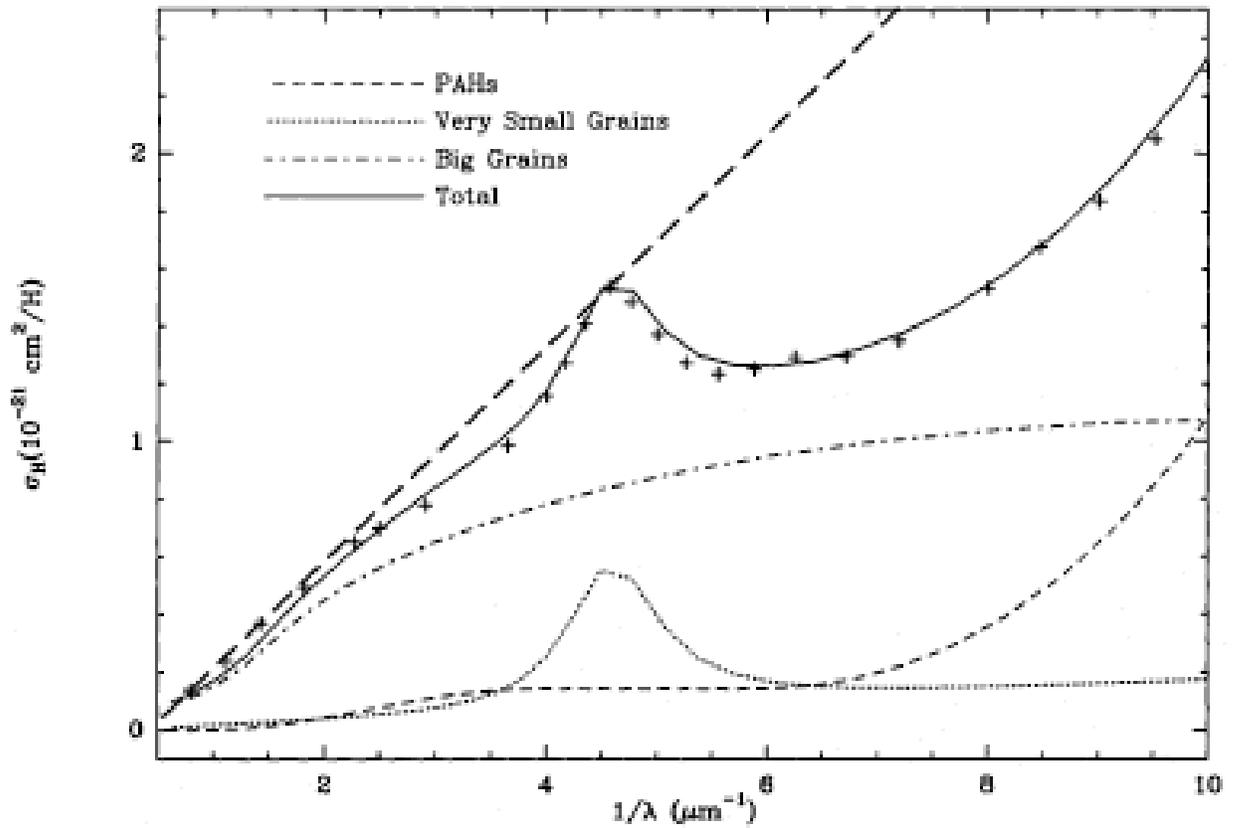}} 
\caption{Seaton's extinction curve (plain line) and the separate 
extinction curves of the large grains, the very small grains (VSG), 
and the poly-aromatic molecules (PAH). From \citet{desert}. I have 
added (dashes) the linear extinction curve.
} 
\label{fig:seaton}
\end{figure*}

\begin{figure*}
\resizebox{!}{0.8\columnwidth}{\includegraphics{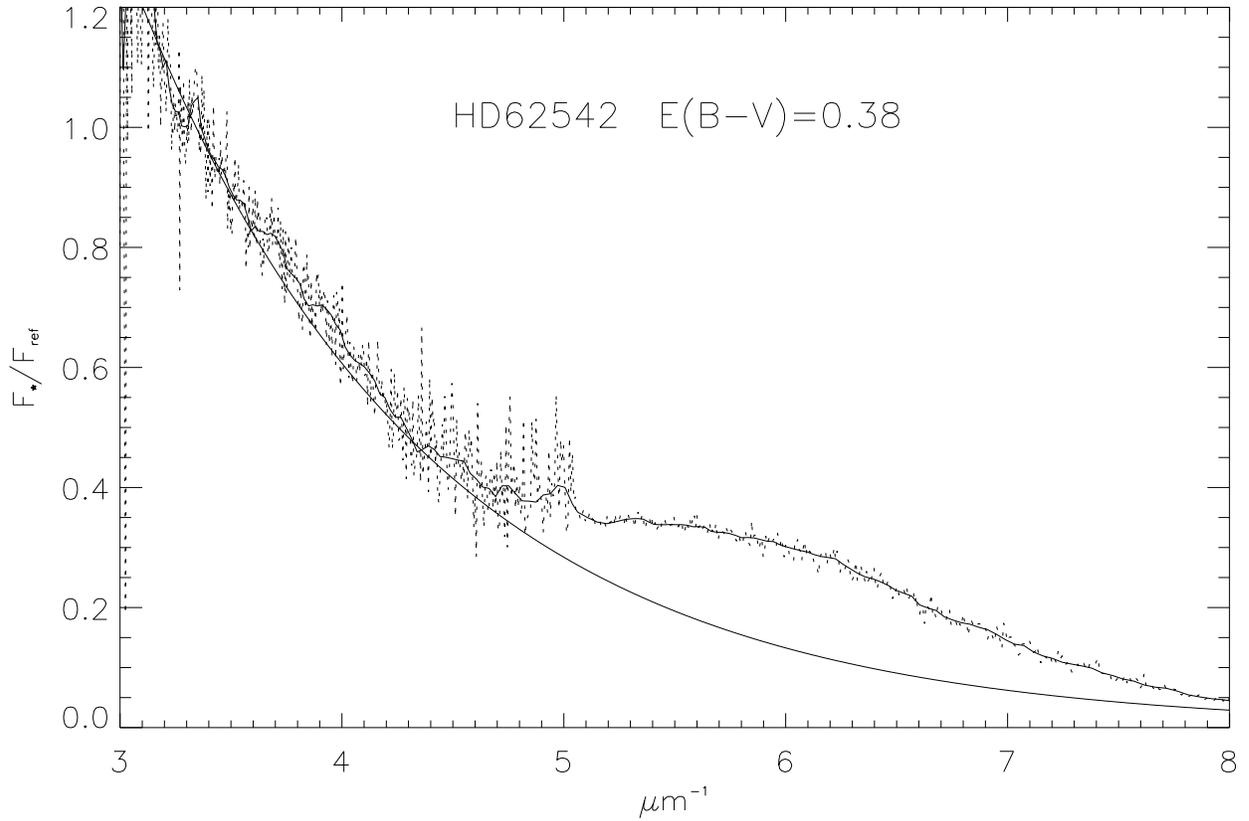}} 
\caption{The spectrum of HD62542 divided by the spectrum of the 
non-reddened reference star HD32630. The visible extinction 
corresponds to the exponential and extends to the bump region. 
The far-UV difference between the two curves clearly appears as an 
excess of light superimposed on the tail of the exponential. This 
excess is attributed to an additional component of scattered light.
} 
\label{fig:hd62542}
\end{figure*}

The observations of section~\ref{obs} are explained if a 
non-negligible contribution of scattered light is added to the observed direct starlight. 
When there is little interstellar matter between the star and the observer 
the contribution of scattered light is negligible: 
we observe the direct starlight only, slightly extinguished, and the observed extinction reflects 
the exact extinction law of starlight by interstellar dust. 
If the column density is increased, scattered light first appears in the far-UV, 
because extinction, hence the number of photons available for 
scattering, increases towards the shortest wavelengths.
Still increasing the reddening, the scattered starlight will merge in 
the visible, provoking the departure of Seaton's curve from the linear 
extinction between $1/\lambda \sim 2.5\mu\rm m ^{-1}$ 
and $1/\lambda \sim 4\mu\rm m ^{-1}$ (figure~\ref{fig:seaton}).

There are many consequences.
Firstly, the extinction curve is a straight line from the near 
infrared to the far-UV.
Secondly, there is a priori non reason to suppose changes of the 
average properties of interstellar dust from one interstellar cloud 
to another:
the law of starlight scattering found for the nebulae for instance is 
the same in the different nebulae.
If, in the near future, spectral observations of nebulae
confirm that this law extends to the visible, we will also 
be able to conclude that the albedo and phase function of interstellar 
grains are wavelength-independent from the near infrared to the far-UV.
Variations of the $R_V=A_{V}/E(B-V)$ parameter are observed for some 
stars and used as a proof of variations of the properties of interstellar dust.
These variations will as well be exlained by the slight modification, due to the scattered starlight, 
of the slope, $2E(B-V)$, of the visible extinction curve of the stars \citep{uv3}.

It also follows that current models of interstellar dust, formed on 
an overly direct interpretation of the particularities of the extinction curve, are questionnable.

There are two practical aspects of this research which merit 
mentionning.
The papers I have published in the two preceding years show the need of 
acquiring stellar spectra on an as large as possible wavelength 
range. 
The simultaneous study of visible and UV extinctions becomes necessary 
to understand and to separate the different contributions of 
direct and scattered starlights.
This necessity of separating direct and scattered starlight 
implies that the current method of using the magnitudes of the spectra rather 
than the spectra themselves is not recommended.
The traditional way would be best if there was direct starlight 
only, but the separation of two additive components is much easier 
from the raw spectra than from their logarithm.
{}
\end{document}